\documentclass[reprint,aps,pre,floatfix]{revtex4-2}
\usepackage[utf8]{inputenc}
\usepackage{amsmath}
\usepackage{graphicx}
\usepackage{physics}
\usepackage[citecolor=black,%
filecolor=black,%
linkcolor=blue,%
urlcolor=black,%
colorlinks=true]{hyperref}
\usepackage{natbib}

\newcommand{\upd}{\text{d}} 

\newcommand{\slfrac}[2]{\left.#1\middle/#2\right.}
\renewcommand{\vec}{\boldsymbol}

\begin{document}
\title{Field theory of enzyme-substrate systems with restricted long-range interactions}
\date{\today}
\author{Fabrizio Olmeda}
\affiliation{Max Planck Institute for the Physics of Complex Systems, Nöthnitzer Strasse 38, Dresden, D-01138, Germany}
\affiliation{Institute of Science and Technology Austria, Am Campus 1, 3400 Klosterneuberg, Austria}
\author{Steffen Rulands}
\email{rulands@lmu.de}
\affiliation{Arnold Sommerfeld Center for Theoretical Physics and Center for NanoScience, Department of Physics, Ludwig-Maximilians-Universität München, Theresienstrasse 37, D-80333 Munich, Germany}
\affiliation{Max Planck Institute for the Physics of Complex Systems, Nöthnitzer Strasse 38, Dresden, D-01138, Germany}

\begin{abstract}
    Enzyme-substrate kinetics form the basis of many biomolecular processes. The interplay between substrate binding and substrate geometry can give rise to long-range interactions between enzyme binding events. Here, we study a general model of enzyme-substrate kinetics with restricted long-range interactions described by an exponent $-\lambda$. We employ a coherent-state path integral and renormalization group approach to calculate the first moment and two-point correlation function of the enzyme-binding profile. We show that starting from an empty substrate the average occupancy follows a power law with an exponent $1/(1-\lambda)$ over time. The correlation function decays algebraically with two distinct spatial regimes characterized by exponents $-\lambda$ on short distances and $-(2/3)(2-\lambda)$ on long distances. The crossover between both regimes scales inversely with the average substrate occupancy. Our work allows to associate experimental measurements of bound enzyme locations with their binding kinetics and the spatial confirmation of the substrate.
\end{abstract}
\maketitle
The binding of enzymes to substrates can catalyze chemical reactions. Enzyme-substrate kinetics therefore form the basis of many biochemical processes. Such kinetics include, for example, the binding of polymerase molecules to the DNA for the transcription of genes to mRNA molecules~\cite{Albe_2002_book}, the binding of oxidase enzymes to the outer membrane of mitochondria for the oxidation of monoamines, or the multi-site phosphorylation of proteins~\cite{Salazar}.

Enzymes can interact in many ways when binding to the substrate. Transcription factors, which control the transcription of genes, often need to bind combinatorically together with other transcription factors in order to initiate transcription~\cite{Reiter:2017aa}. Some enzymes seem to slide diffusively along the substrate, such as the methyl transferase DNMT1~\cite{VILKAITIS200564}. In many biologically relevant scenarios the substrate undergoes confirmational changes when bound by an enzyme~\cite{KOSHLAND:1963aa}. This is, for example, used to increase specificity of enzyme binding in a kinetic proof-reading scheme~\cite{proofreading}. Many enzymes that bind to the DNA, including chromatin modifiers, cause confirmation changes thereof, most notably the compaction and decompaction of the DNA~\cite{Dekker2015}. Positions that are far apart measured along the DNA sequence might then be close in physical space. Due to the feedback of enzyme binding with chromatin confirmation this give rise to effective long-range interactions between binding events.

Experimentally, novel technologies in molecular biology allow quantifying the binding locations of enzymes to substrates like membranes or the DNA in a static manner. These technologies include super-resolution microscopy such as DNA Paint~\cite{Schnitzbauer:2017aa}, which can image a wide range of enzyme binding events with the resolution of individual enzymes~\cite{Reinhardt:2023aa}. For one-dimensional substrates, single-cell sequencing technologies allow measuring the consequences of enzyme binding, such as chemical modifications of the DNA or histone tails, with the resolution of single base pairs and in individual cells~\cite{Clark:2018aa,Gawad:2016aa}. A theoretical prediction about the relation between the statistics of bound positions on the substrate and the kinetics of enzyme binding would set the basis for concluding the enzyme kinetics and substrate confirmation from such experiments.

In thermal equilibrium, the binding and unbinding of enzymes is strictly constrained by the condition of detailed balance. In this case, the equilibrium enzyme binding profile is predicted by the Gibbs free energy \cite{langmuir_2018_1429050}. In the context of cell biology, enzyme-substrate kinetics is often out-of-equilibrium. Prominent examples of such kinetics are chromatin modifies, which catalyze the irreversible deposition of epigenetic marks on the DNA or histone tails and may lead to confirmational changes of the DNA. In these cases, a description in terms of thermodynamic potentials is not feasible and the theory of stochastic processes provides a general framework for describing non-equilibrium enzyme-substrate kinetics~\cite{gardiner2004handbook,reuveni2014role,Morelli2011,JULICHER19981169}.

Biological substrates and enzymes are often highly nonlinear and the spatial dimensions cannot be neglected~\cite{Li2009,JULICHER19981169,COPPEY20041640,Mirny_2009,PhysRevLett.110.208104,Gupta2017}. In these cases, standard approximation schemes of master equations and Langevin equations do not exist or are difficult to use. In these cases, path-integral representations of stochastic processes provide a versatile and and powerful framework for studying stochastic processes. These include the Martin-Siggia-Rose-Janssen-de Dominicis functional integral representation of stochastic differential equations and the coherent-state path integral representation of master equations~\cite{Altland}. In both cases, expectation values of observables can be expressed in terms of deterministic path integrals over a pair of conjugated fields. Path-integral representations allow for the application of powerful theoretical tools, such as perturbation theory and renormalization group theory~\cite{ZinnJustin}, in the context of stochastic processes.

In this work, we develop a theoretical framework of the out-of-equilibrium enzyme-binding kinetics on substrates. Specifically, we use a coherent-state path-integral approach in combination with renormalization-group calculations to obtain the moments of the distribution of enzymes bound on the substrate. We show that the first moment, the fraction of bound sites, increases according to a power-law with an exponent depending on the decay exponent of long-range interactions. The two-point correlation function decays in two temporal regimes characterized by different exponents. For short distances, it is dominated by active feedback between enzyme binding events and for long distances by conservative noise.

\begin{figure}[tb]
    \centering
    \includegraphics[width=0.85\linewidth]{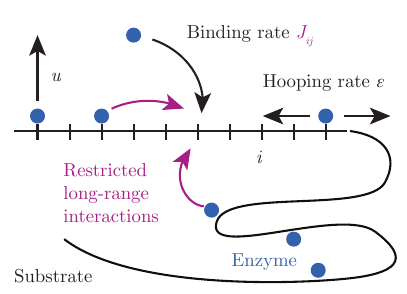}
    \caption{Schematic depicting the stochastic enzyme-substrate model. In this model, enzymes bind to a substrate with a binding rate that depends on the distance to the nearest bound sites. In the context of cell biology, substrates like DNA have dynamic geometries, such that local interactions in space lead to long-range interactions along the linear lattice.}
    \label{fig:enter-label}
\end{figure}

\section{Model definition}
\label{seq:modeldefinition}
We consider a general model for enzyme-substrate kinetics. In this model, the substrate is represented by a one-dimensional lattice of size $N$. On this lattice, each position is a potential binding site indexed by $i\in\{1,\ldots,N\}$. Enzymes can bind to and unbind from lattice sites with rates that depend on the positions of other bound enzymes. This dependence of the binding rates is encoded in an interaction kernel $J_{ij}$ which gives the contribution of a bound enzyme at position $j$ to the enzyme binding rate at position $i$.

Since in biological systems infinite-range interactions are implausible we here restrict these interactions to the nearest bound sites. Specifically, if the nearest bound sites are at positions $j_1$ and $j_2$ the interaction kernel is $J_{i,j_1} + J_{i,j_2}$, where the two terms correspond to the contribution from the left and right bound neighbour, respectively. Such interactions are in the literature known as restricted long-range interactions~\cite{ginelli2005directed,Ginelli,Hinrichsen}. We further only consider translationally invariant kernels of the form $J(|i-j|)$. The unbinding rate of the enzyme is constant and we denote it by $u$. In equilibrium, the binding and unbinding rates would be related by the condition of detailed balance. Here, since many enzymatic processes are microscopically irreversible through the conversion of ATP to ADP, we do not make this assumption and take the unbinding rate to be independent. We further allow enzymes to perform random walks along the substrate with a rate $\epsilon$. 

With this, the state of the system is described by a random vector $\vec{\sigma}$ with entries representing enzyme occupancy: $\sigma_i = 1$ if site $i$ is bound and $\sigma_i = 0$ otherwise. The time evolution of the probability of finding a given binding profile $\vec{\sigma}$ at a time $t$ then follows a master equation of the form
\begin{widetext}
\begin{align}
\label{eq:master_eq}
   \frac{\partial P(\vec{\sigma})}{\partial t} = &
    \sum_{i=1}^{N} \sum_{l=i+1}^{N} J_{i,i+l}\left(\prod_{j=1}^{l-1}\bar{\sigma}_{i+j}\right) \sigma_l [\sigma_i  P(\vec{\bar{\sigma}}_{i}) - \bar{\sigma_i}P(\vec{\sigma})] + u \sum_{i=1}^{N}  \sigma_i \left[P(\vec{\bar{\sigma}}_{i}) - P(\vec{\sigma}) \right]\\
    &+ \epsilon \sum_{i=1}^{N} \left[\bar{\sigma}_{i}\sigma_{i+1}P(\vec{\bar{\sigma}}_{i}) - \sigma_i\bar{\sigma}_{i+1}P(\vec{\bar{\sigma}}_{i+1}) \right] 
   +\text{l. n. n.} +\text{l. h.} \, .
\end{align}
\end{widetext}
Here, we introduced a notation for the vector $\vec{\bar{\sigma}}_i$ which has the same elements as $\vec{\sigma}$, but the value of $\sigma_i$ at position $i$ is replaced by $1-\sigma_{i}$. Similarly, we define $\bar{\sigma}_i = 1-\sigma_i$. In Eq.~\eqref{eq:master_eq} the first term describes the contribution of interactions with the right nearest neighbor to the enzyme-binding rate. The contribution from the left nearest neighbor takes the same form and is here abbreviated by l.~n.~n.. The second terms describes unbinding events and the third term enzyme hopping to to the right. Hopping events to the left again take the same form and are abbreviated by l.~h..

\section{Coherent-state path integral formulation of the master equation}
Having defined a general model for the kinetics of enzyme-substrate binding in terms of Eq.~\eqref{eq:master_eq}, we now investigate the non-stationary distribution $P(\vec{\sigma},t)$ in terms of its moments. To this end, we employ a coherent-state path integral formulation of the master equation, Eq. \eqref{eq:master_eq}~\cite{Altland}. Taking the semi-classical limit we will compare the prediction for the first and higher-order moments to numerical simulations. 

To define the path-integral representation we first introduce a Fock space, in which we represent lattice configurations in bra-ket notation. We define creation operators $a_i^{\dagger}$which formally represent the binding of enzymes to site $i$. With these, we can express a given state $\ket{\sigma} = \ket{\sigma_1,\ldots,\sigma_N}$ as the repeated application of creation operators on the empty lattice, $\ket{0}$,
\begin{equation}
    \ket{\sigma} = a_1^{\dagger \sigma_1}...a_N^{\dagger \sigma_N}\ket{0}\, .
\end{equation}
We also define annihilation operators, $a_i$, which represent the unbinding of enzymes from a given site $i$. The annihilation operators are conjugate to the creation operators, such that both act on a given lattice configuration as follows,
\begin{equation}
\begin{split}
a_i^{\dagger}\ket{\sigma} &= \ket{\sigma_i+1}\, ,\\
a_i\ket{\sigma} &= \sigma_i \ket{\sigma_i-1}\, .
\end{split}
\end{equation}
With these definitions, the creation and annihilation operators follow standard commutation rules,
\begin{equation}
[a_i,a^{\dagger}_i] =1.
\end{equation}
We can now write the probability distribution in Fock space formally as \cite{Cardy},
\begin{equation}
    |P(t) \rangle  = \sum_{\vec{\sigma}} P(\vec{\sigma},t)a_1^{\dagger \sigma_1}...a_N^{\dagger \sigma_N}|0\rangle\, .
\end{equation}
Using this notation, we can formally rewrite the master equation in terms of the creation and annihilation operators,
\begin{equation}
\partial_t |P(t) \rangle= - H |P(t) \rangle\, ,
\end{equation}
with a "Hamiltonian" $H$ defined as
\begin{align}
 H   = & \sum_{i=1}^{N} \sum_{l=1}^{N-i} J_{i,i+l} \prod_{j=1}^{l-1} \Big[ (1-a^{\dagger}_{i+j}a_{i+j}) \\
    &\quad\quad \cdot a^{\dagger}_{i+l}  a_{i+l}(a^{\dagger}_{i}\hat{\delta}_{\sigma_{i,0}} - \hat{\delta}_{\sigma_{i,0}} ) \Big] \\
     + &\sum_{i=1}^{N-1}\epsilon(1-a_i a^\dagger_{i+1} )\hat\delta_{\sigma_i,1}\delta_{\sigma_{i+1},0} \\
    & + \text{l. n. n.} + \text{l. h.}\, .
\end{align}
The operators, $\hat\delta_{\sigma_{i},k}$ are equal to the identity if a site $i$  is occupied by $k$ enzymes and 0 otherwise. These operators enforce that only a single enzyme can be bound at a given site. Although we will refer to $H$ as a Hamiltonian, $H$ does not represent an energy as it is not necessarily hermitian unless detailed balance is fulfilled.  

With this, we can write the expectation value of any observable $A(\vec{\sigma},t)$ as
\begin{equation}
\langle A (\vec{\sigma},t)\rangle = \sum_{\vec{\sigma}} A(\vec{\sigma}) P(\vec{\sigma},t)\, .
\end{equation}
We can rewrite this equation as
\begin{equation}
\langle A (\vec{\sigma},t)\rangle = \bra{0} \prod_i\sigma_i e^{a_i} A(\vec{\sigma}) \ket{P(t)}\, ,
\end{equation}
where we introduce a coherent state basis $\bra{0}e^a $, which has the property to be the left eigenstate of the creation operator $a^{\dagger}$,
\begin{equation}
\bra{0}e^a a^{\dagger} = \sum_{n=1}^{\infty} \frac{\bra{0}}{n!} a^{n} a^{\dagger}= \bra{0}e^a.
\end{equation}

Having defined all the rules of the operators (cf. \hyperref[sec:AppendixA]{Appendix \ref{sec:AppendixA}}) we now proceed to derive the path-integral representation of the master equation. To this end, we define a continuous field, $\phi_i(t)$, giving the local density of bound sites at position $i$. We then use the decomposition of 1,
\begin{equation}
    1=\int \upd\phi_i\, \upd\hat{\phi}_i\, e^{-\hat{\phi}_i\phi_i}e^{\phi_i a_i^{\dagger}}\ket{0}\bra{0}e^{\hat{\phi}_i a_i}\, ,
\end{equation}
to derive a path-integral representation of the master equation following standard steps~\cite{Cardy, Tauber}. This decomposition introduces a second field, $\hat\phi$, which is called response field or conjugated field. In this path-integral representation the moments of the distribution $P(\vec{\phi},t)$ can be expressed in terms derivatives of the generating functional \cite{doi1976second,peliti1985path}, 
\begin{equation}
\label{eq:correlationFT}
Z[\vec{h},\phi,\hat{\phi}]= \int \mathcal{D}[\vec{h},\phi,\hat{\phi}]   e^{-S[\phi,\hat \phi] + \int \upd t \sum_i\left[h_i\phi_i + \hat{h}_i\hat{\phi_i} \right]}\, ,
\end{equation}
where we used the notation $\mathcal{D}[\vec{h},\phi,\hat{\phi}]$ to denote the integrals over all positions $\prod_i\upd\phi_i\upd\hat\phi_i\upd h_i$. The argument in the exponential is an action of the form, 
\begin{equation}
S[\hat{\phi},\phi] = - \sum_{i}\phi_{i}(t_f) + \int_{0}^{t_f} \mathrm{d}t \sum_{i}\left(\hat{\phi}_{i} \partial_t \phi_i
+ H_i[\hat{\phi},\phi]\right)\, .
\end{equation}
with Hamiltonian,
\begin{equation}
    \label{hamiltonian}
H_i =(1-\hat{\phi}_i)  e^{-\hat{\phi}_i \phi_i}  \sum_{l=1}^{N-i}\prod_{j=1}^{l-1}  J_{i,i+l}(l) \hat{\phi}_{i+l}  \phi_{i+l} 
(1- \phi_{i+j})   + \text{o. t.}\, ,
\end{equation}
where we summarized the hopping and left-nearest-neighbor interactions by o.~t..

In order to evaluate the path integral, we define how the $\hat{\delta}$ operators in $H$ act on the coherent-state basis by following the rules in Ref.~\cite{Wijland},
\begin{equation}
\label{eq:exclusion}
\begin{split}
&\bra{\phi}   a_i^{\dagger}\hat{\delta}_{{n_i},m_i} \ket{\phi_i} = \frac{1}{m_i!}  \hat{\phi_i} \left( \hat{\phi_i}  \phi_i  \right)^{m_i} e^{-\phi_i\hat{\phi_i}}\, ,\\
&\bra{\phi}   a_i\hat{\delta}_{n_i,m_i}  \ket{\phi} = \frac{1}{(m_i-1)!} \phi_i\left( \hat{\phi_i}  \phi_i  \right)^{m-1} e^{-\phi_i\hat{\phi_i}}\, ,\\
&\bra{\phi} \hat{\delta}_{n_i,m_i} \ket{\phi}  =\frac{1}{m_i!}   \left( \hat{\phi_i}  \phi_i  \right)^{m_i} e^{-\phi_i\hat{\phi_i}}\, .
\end{split}
\end{equation}
Many different approaches have been proposed to deal with the particle exclusion enforced by the $\hat\delta$ operator. Examples for this are fermionic field theories \cite{brunel2000fermionic} or more recent approaches using negative rates \cite{nekovar2016field}. Here, we will use the hard-bosonic path-integral approach \cite{Wijland}. The benefit of using this formalism is that it gives a characteristic length scale, which distinguishes different spatial regimes of the correlation function and which will be important for its calculation.

\subsection{Semiclassical limit of the field theory}
In order to obtain the moments of $P(\vec{\sigma},t)$, we now derive the semiclassical solution of the field theory. To this end, we will evaluate the path integral in Eq.~\eqref{eq:correlationFT} for fields that make the action $S$ extremal. This limit will be shown to be valid if the average enzyme occupancy is small.

Solutions to this model with \emph{local} interactions are known or can easily be obtained using the methods we describe below. 
Here, we study interactions that are motivated by the interplay between enzyme binding and geometric changes of the substrate. For example, enzyme binding could compact the substrate, such that sites far away on the lattice are in close proximity in real space. Such interactions are therefore long-ranged in nature and we consider a general class of non-local interaction kernels of the form $J_{i,i+l}=1/l^\lambda$. 

In a first step, we rewrite the Hamiltonian, Eq.~\eqref{hamiltonian}, in continuous space with coordinate $s$. Upon introducing a spatial discretization $\sum_i \Delta s \rightarrow \int ds$, where $\Delta s$ is the lattice spacing and $s$ a continuous coordinate, the Hamiltonian in the action, Eq.\eqref{hamiltonian} becomes,
\begin{equation}
\label{eq:H_nofieldchange}
\begin{split}
&H[\hat{\phi},\phi] =  J(1-\hat{\phi} (s)) e^{-\phi \hat{\phi}} \cdot \\
&\cdot\Big[ \int_{0}^{N-s} \mathrm{d}y \frac{\hat{\phi}(s+y) \phi(s+y)}{y^{\lambda}} 
e^{-\int_{z=0}^{y} \mathrm{d}z \hat{\phi}(s+z)  \phi(s+z)} \\ 
& +\int_{0}^{s} \mathrm{d}y \frac{\hat{\phi}(s-y) \phi(s-y)}{y^{\lambda}} 
e^{-\int_{z=0}^{y} \mathrm{d}z \hat{\phi}(s-z)  \phi(s-z)} \Big]\, ,
\end{split}
\end{equation}
Here we choose the binding rates $J_{ij}$ to be equal throughout the lattice, i.e. we do not consider disorder. As the integrals in Eq.~\eqref{eq:H_nofieldchange} diverge for $\lambda \geq 1$,  we henceforth require that $\lambda < 1$. 

Expanding Eq.~\eqref{eq:H_nofieldchange} to first order in the exponentials, to the second order in the terms $\slfrac{\hat{\phi}(s \pm y) \phi(s \pm y)}{y^{\lambda}}$ and extending the upper limit of integration to infinity the terms containing the integrals simplify to 
\begin{equation}
    \left[ 2 \left(\hat{\phi}\phi\right)^{\lambda} + \left(\phi \hat{\phi}\right)^{\lambda-3}\left(2-3 \lambda + \lambda^{2}\right)\frac{\partial^2 \left( \phi \hat{\phi}\right)}{\partial s^2} \right].
\end{equation}

We now perform a semi-classical approximation of the generating functional by employing a saddle-note bifurcation of the path integral. To this end, we minimize the action, $\slfrac{\delta S}{\delta \hat{\phi(s)}}|_{\hat{\phi}(s) = 1} =0 $,  while setting $\hat{\phi}(s) = 1$ to ensure probability conservation (cf.  Ref. \cite{Tauber}). With this, we obtain a partial differential equation describing the time evolution of the enzyme-binding profile along the substrate, $\phi(s)$, which for small values of $\phi(s)$ is, 
\begin{equation}
\label{eq:Langevin}
\begin{split}
\frac{\partial \phi (s)}{\partial \tilde{t}} = &e^{-\phi(s)}\left\{ \phi(s)^{\lambda} + \left[\mu \phi(s)^{\lambda -3} +  \epsilon\right]\partial^2_{s} \phi(s)\right\} \\
&+ e^{-\phi(s)}\eta(s,t) \, .
\end{split}
\end{equation}
Here, we rescaled time according to $\tilde{t} = t 2J \Gamma(1-\lambda)$ and defined $\mu =\slfrac{(2-3 \lambda + \lambda^{2})}{2}>0$. We reuse the symbol $\epsilon$ for the diffusion constant. The factor $e^{-\phi(s)}$ is a direct consequence of the site restriction, which intuitively has the role of  preventing unbounded growth.
e
In the hard-boson path-integral representation, Eq.~\eqref{eq:H_nofieldchange}, the term $\exp\left[-\int_{z=0}^{y} \mathrm{d}z \hat{\phi}(s+z)  \phi(s+z)\right]$ in the Hamiltonian is approximated by an exponential of the space-averaged fields, $\exp\left[-y\langle\hat{\phi}(s)  \phi(s) \rangle\right]$ if the fields are slowly varying. This defines an effective exponential cutoff of the interactions at a characteristic length $\slfrac{1}{y^{\lambda}}$. Because contributions from sites separated by a long distance are effectively suppressed, this justifies the expansion of the limit in Eq.~\eqref{eq:H_nofieldchange} to $\infty$. 

\begin{figure}
    \centering
\includegraphics[width=0.95\linewidth]{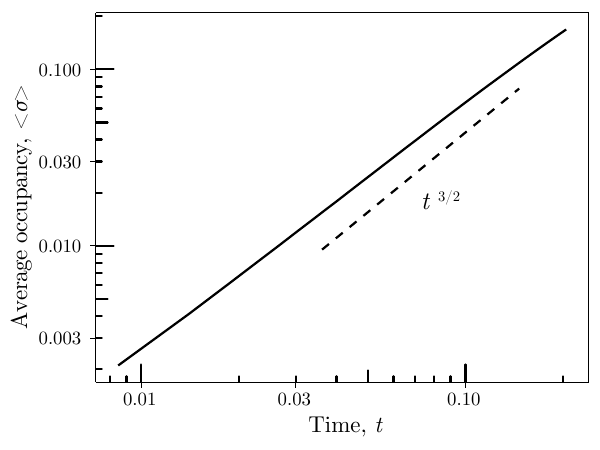}
   \caption{Time evolution of the average enzyme occupancy obtained from stochastic simulations of the master equation \eqref{eq:master_eq} for $\lambda =1/3$ and $10^7$ lattice sites (solid line). The simulation was initialized with For early times, the theory predicts and increase following a power law with an exponent of $3/2$ which is indicated by the dashed line.}
    \label{fig:Powerlawgrowth}
\end{figure}

The first spatially homogeneous moment, $\phi_0(t)$, i.e. the average number of bound enzymes, $\langle \sigma \rangle$, is the solution of Eq.~\eqref{eq:Langevin} when neglecting diffusion. With an almost empty substrate as an initial condition, the term $e^{-\phi}\phi^{\lambda}$ on the right hand side of Eq.~{\eqref{eq:Langevin}} is to first order well approximated by $\phi^\lambda$. Therefore, at short times, the first moment follows a power-law,
\begin{equation}
    \langle \sigma \rangle \sim t^{\frac{1}{1-\lambda}} \, .
\end{equation}
 This is the first result of this manuscript. In Fig.~\hyperref[fig:Powerlawgrowth]{\ref{fig:Powerlawgrowth}} we compare the analytical solution to stochastic simulations using Gillespie's algorithm~\cite{Gillespie}. See \hyperref[sec:numericalsimul]{Appendix \ref{sec:numericalsimul}} for details of the numerical implementation. 

 At large times, site exclusion becomes dominant and the solution approaches a steady-state solution, $\phi^*$, logarithmically. The steady-state concentration of bound enzymes is determined by a balance between binding and unbinding events and it is given by
\begin{equation}
     \phi^*=(1-\lambda) W\left(\frac{\bar{u}^{\frac{1}{\lambda-1}}}{1-\lambda}\right)\, ,
 \end{equation}
 where $W$ is the Lambert $W$ function and $\bar{u} = u/[2J \Gamma(1-\lambda)]$ is the relative strength of unbinding and interactions. Therefore, if the coupling, $J$, is much larger then $u$, the steady-state concentration of bound enzymes decreases with $\bar{u}^{-1/(1-\lambda)}$.  In the limit that interactions are much weaker compared to degradation the steady-state concentration in creases logaithmically, $\phi^*\sim\ln\left[\bar{u}^{-1/(1-\lambda)}\right]$.
 
 Taken together, starting from an almost empty substrate, the average number of occupied binding sites increases following a power law up to a point where higher-order interactions due to site-exclusion become dominant. As an example, for enzymes binding to a fully compacted, space-filling  polymer, $\lambda$ takes a value of $1/3$ such that the number of bound sites increases with an exponent of $3/2$.

As a remark, at the level of a mean-field description, we can map the enzyme-substrate model to a Smoluchowski equation \cite{krapivsky2010kinetic} describing the size of domains where all sites are bound. In this view, a binding event between two already occupied sites is equivalent to a fragmentation event splitting the domain into two smaller intervals. Restricting our attention to the first moment, we can rewrite our model in the mean-field limit as a fragmentation equation,
\begin{equation}
\label{fragmentation}
\frac{\partial c(x)}{\partial t} =   \int_{0}^{\infty} \mathrm{d}y\, J(x,y-x)c(y) - c(x) \int_{0}^{x}\mathrm{d}y \,J(y,x-y) \, ,
\end{equation}
where $c(x,t)$ is the number of domains of size $x$ at time $t$ and we omitted the time-dependency for a shorter notation. For the chosen kernel,  $J(x,y) =  \slfrac{1}{x^{\lambda}} +\slfrac{1}{y^{\lambda}}$, and after defining the moments as $M_{\alpha} = \int \mathrm{d}x\,  c(x) x^{\alpha}$, we obtain $M_{\alpha + \lambda + 1} \sim t^{-\slfrac{(\alpha + \lambda )}{(\lambda +1)}}$. The average occupancy, which is the zeroth moment, scales as $\langle\sigma\rangle \sim t^{ \slfrac{1}{(\lambda - 1)} }$.  This equation  gives the same time evolution of the number of enzymes bound to the substrate as the path-integral methods.

\subsection{Correlation functions}
To compute correlation functions and scaling exponents we need to take the derivative of the generating function \eqref{eq:correlationFT}. However, such an approach is not feasible for three reasons. 

The first reason is that we would need to expand the action and this attempt fails due to the presence of a mass term in the field theory. Therefore, the field theory is not scale-invariant. In order to overcome this problem, we could try to do a change of variables in the action Eq.\eqref{eq:H_nofieldchange}, and make a Hopf-Cole transformation around a dynamical mean field solution, such that in that reference frame the theory is massless. Even though the theory will be renormalizable, the exponent we get from such a calculation does not agree with the result of numerical simulations.  The second reason why such an approach fails is that we would need to consider perturbations of any order in the field theory and a one-loop calculation would not be sufficient~\cite{JannssenWijland}. The third and main reason for the failure of this approach is that, after expanding around a base state, integrals of the form in Eq.~\eqref{eq:H_nofieldchange} cannot be approximated without losing the length scale, $1/\langle \phi\rangle$, associated with the effective cutoff of the interactions.

The key insight to calculate the correlation function is that Eq.~\eqref{eq:H_nofieldchange} gives rise to two spatial regimes: at short distances, interactions are long-ranged following a power law decay with exponent $1/\lambda$ while for distances much larger than $1/\langle \phi \rangle$ interactions are effectively screened. In the following, we will therefore derive the correlation function separately for these two regimes using renormalization group methods and perturbation theory. We will then confirm these results with numerical simulations.

\subsubsection{Short-distance regime}
\label{sec:shortdistanceregime}
 
To describe the short-distance behavior of correlation functions, we start from the action, Eq.~\eqref{eq:H_nofieldchange}, and, after taking the semi-classical approximation, we expand it to first order in  the fields and their derivatives,
\begin{equation}
\begin{split}
     \label{perturbationlangevin}
&\partial_t \phi(s,t)   = 
\int_0^{s} dy\, \phi(y) |s-y|^{-\lambda}e^{- \int_{z=0}^{s-y}dz\,\phi(z)}+ \\
&\int_0^{s} dy\, \partial_y \phi(y)|s-y|^{1-\lambda}e^{- \int_{z=0}^{s-y}dz\,\phi(z)} \, ,
\end{split}
\end{equation}
where $s$ is again a continuous coordinate giving the position on the substrate. Here, for notation clarity, we omitted the noise terms, hopping terms, and integrals of the same form describing interactions with the right nearest bound site but we will consider them at the end of the derivation. 

The interaction kernel Eq.~\eqref{eq:H_nofieldchange} has the form $|s-y|^{-\lambda} e^{- \int_{z=0}^{s-y}dz\,\phi(z)}$, with $\lambda <1$. By considering a perturbation $h(s,t)$  around the mean-field solution, $\phi_0(t)$, i.e. $\phi(s,t) = \phi_0(t) + h(s,t)$, Eq.~\eqref{perturbationlangevin} can be expressed to first order as
\begin{align}
     \partial_t \phi_0  +  \partial_t h  & =  e^{-\phi_0}\int_0^{s} \upd y\, \phi_0  |s-y|^{-\lambda}\left[1- \int_{0}^{s-y}\upd z\, h(z)\right] \nonumber\\ 
   &+e^{-\phi_0}\int_0^{s} \upd y\, h(y)   |s-y|^{-\lambda}\left[1- \int_{0}^{s-y} \upd z\, h(z)\right]\nonumber\\
   &+\text{h.o.t}\, .
\end{align}
The first two terms of the integral on the right-hand side cancel with the first one on the left-hand side, which follows from the dynamical mean field solution, Eq.~\eqref{eq:Langevin}. Making a change of variables, $w = z+y$,  we then obtain
\begin{equation}
 \label{eq:short_distance_langevin}
 \begin{split}
     \partial_t h  = &e^{-\phi_0}\int_0^{s} \upd y\, h(y)   |s-y|^{-\lambda}  \\
     -&e^{-\phi_0} \int_0^{s} \upd y \int_{y}^{s} \upd w\, h(y)   |s-y|^{-\lambda} h(w-y) + \xi(s,t)\, .
 \end{split}{}
\end{equation}
The second term on the right-hand side of Eq.~\eqref{eq:short_distance_langevin} is the convolution of a fractional integral of a function and the function itself.

The noise term, $\xi(s,t)$, is Gaussian white noise. Its correlations can be  derived from the field theory, Eq.~\eqref{eq:H_nofieldchange}, by identifying terms that are proportional to $\hat{\phi}^2$. In the expansion of the action these terms comprise both non-conservative noise, which means that they are proportional to $\hat{\phi}^2$, and conservative noise, which means that they are proportional to $\hat{\phi}^2 \partial_s^2\phi$.  The perturbation $h(s,t)$ thus has both conservative and non-conservative components, such that we make the following ansatz for the noise correlations:
\begin{equation}
    \langle \xi(s,t) \xi(s',t') \rangle = \delta(t-t')(2\Gamma_{NC}- 2\Gamma_C \partial_s^2)  \delta(s-s').  
\end{equation}
$\Gamma_C$ and $\Gamma_{NC}$ are the noise strengths for conservative and non-conservative noise, respectively, which are functions of $J$ only. We omit their dependency on the model parameters as this dependence does not influence the exponents of correlation functions. 

Taken together, as the fractional integral scales in Fourier space as $q^{\lambda-1}$, a spatio-temporal perturbation in Fourier space follows a nonlinear Langevin equation of the form
\begin{equation}
\begin{split}
    \partial_t h(q,t)   = & \left( e^{-\phi_0(t)} q^{\lambda-1}- q^2\right)h(q,t) \\
  & -  q^{\lambda-1}e^{-\phi_0(t)}h(q,t)^2 + \xi(q,t)\, .
\end{split}
\end{equation}
From the dynamical mean-field solution of the first moment, Eq.~\eqref{eq:Langevin}, we obtain, up to a prefactor, that $e^{-\phi_0} = e^{\left[{-t^{\slfrac{1}{(1-\lambda)}}}\right]}$. In the frequency domain, we hence obtain at small times, $t \rightarrow 0$ or $\omega \to \infty$, that
\begin{equation}
\label{eq:h_fourier}
    G_0^{-1} h(q,\omega)   =  -  q^{\lambda-1}h(q,\omega)^2 + \xi(q,\omega)\, ,
\end{equation}
where we have defined the inverse bare propagator as $G_0^{-1} = i\omega  + q^2 -  q^{\lambda-1}$. From the above equation, the bare correlations of $h$ are then
\begin{equation}
\label{eq:bare_correlator}
C_0 =\left( 2\Gamma_{NC}+ 2\Gamma_C q^2\right)|G_0|^2\, .
\end{equation}

Eq.~\eqref{eq:h_fourier} is of second order and so admits an exact solution
from which we calculate the two-point correlation function, $\langle h(q,\omega)h(q',\omega') \rangle$, where the average $\langle \ldots \rangle $ is performed over the noise.

As we are considering the short-distance regime, we  keep only leading-order in $q$ and perform the inverse Fourier transform.  With this, we find that  correlation functions asymptotically approach a power law,
\begin{equation}
\label{eq:correlation_shortrange}
 \langle h(s,t)h(s',t) \rangle =  \frac{ \text{cos}(\pi \lambda) \left(\abs{s-s'}^2 \Gamma_C + \Gamma_{NC} \lambda(1 + \lambda)\right)}{\Gamma(\lambda)\abs{s-s'}^{2 + \lambda} } 
\, .
\end{equation}
We therefore obtain for the scaling of the correlation function that
\begin{equation}
    \langle h(s,t)h(s',t) \rangle \sim \abs{s-s'}^{-\lambda}\, ,
\end{equation}
as the conservative noise ($\sim q^2$) is leading with respect to non-conservative noise ($\sim q^0$) for short-distances.
This result is intuitive as the spatial correlations are proportional to the exponent of the scale-free kernel $J$. This derivation is valid for low values of the average occupancy $\phi_0$. We expect that for larger values of $\phi_0$ higher-order corrections become relevant.

\subsubsection{Long-distance regime}
From the bare propagator and correlator in Eq \eqref{eq:bare_correlator} we notice that, at large distances, second-order spatial derivatives dominate correlations, and due to non-conservative noise we expect correlation functions to be described by different exponents compared to the short-distance regime, where conservative noise was the leading contribution in the scaling, Eq.~\eqref{eq:correlation_shortrange}. The mean-field exponents can be estimated by dimensional analysis \cite{Tauber}, where we compute the scaling of the parameters in the action in order to make it dimensionless. Following these arguments, we find from Eq.~\eqref{eq:H_nofieldchange} that the value of the critical exponent of the correlation in the long-distance regime should be $\chi = \slfrac{-(1+d+\lambda)}{3}$. This follows from considering higher-order non-linearities and it is the correct exponent for the long tail of the correlation function.  In the following, we will derive the results of this scaling argument rigorously with the use of dynamical renormalization group methods.

We begin with Eq.~\eqref{perturbationlangevin}. After linearization it becomes,
\begin{align}{}
\label{eq:pde_linear_long_distance}
        \partial_t h(s)  & =  \partial_s^2 h(s)+ \int_0^{s} \upd y h(y)   |s-y|^{-\lambda}   \\  
       &- \int_0^{s} \upd y h(y) |s-y|^{1-\lambda}\left[h(s)-(s-y)\frac{1}{2}\partial_s h(s)\right]\\
       &+ \xi(s,t)\, ,
\end{align}
where we used that $\int_a^b \upd s\, f(s)  \approx (b-a)(f(a)+f(b))/2$.
Before proceeding with a renormalization procedure, we note that the non-linearities of order $h^2$ in Eq.~\eqref{eq:pde_linear_long_distance}  can be relevant in this regime. 

 \begin{figure}
    \centering
    \includegraphics[width=0.95\linewidth]{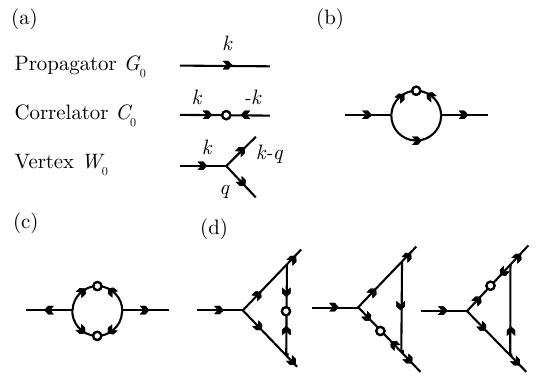}
    \caption{(a) Overview of diagrammatical elements and diagrams contributing to the renormalization of the (b) propagator, (c) correlator and (d) vertex functions. Every intersection corresponds to an integration over the wave-vectors.}
    \label{fig:diagrams}
\end{figure}

Because we approximated to linear order when obtaining Eq.  \eqref{eq:pde_linear_long_distance} we must keep all the other quadratic terms in the field theory, Eq.~\eqref{eq:H_nofieldchange}. There is only one quadratic term which is,
\begin{equation}
\label{eq:extraterm}
\left(\hat{\phi} -1\right) \phi\hat{\phi}
 \int_{0}^{s} \mathrm{d}y \frac{\hat{\phi}(s-y) \phi(s-y)}{y^{\lambda}} 
e^{-\int_{0}^{y} \mathrm{d}z \hat{\phi}(s-z)  \phi(s-z)}.
\end{equation}
However, after functional minimization of the action it cancels out with the term involving interactions with the right nearest neighbor.  This is not surprising, because the model does not allow for field theories involving  terms that break the space-reversal symmetry $s\rightarrow -s$. Taken together, by repeating the same calculations as before and including the additional term in Eq.~\eqref{eq:extraterm} we obtain
\begin{align}
 \partial_t h(s)  & =  \partial_s^2 h(s)+ \int_0^{s} \upd y\, h(y)   |s-y|^{-\lambda} \\
  &   +  \frac{1}{2}\int_0^{s} \upd y\, h(y) |s-y|^{2-\lambda} \partial_s h(s)+ \xi(s,t)\,.
\end{align}
Considering both right and left nearest-neighbor interactions, the advective terms cancel out, again by the necessity to obey left-right symmetry. We then include the next highest-order term,
\begin{equation}
\label{eq:langevin_higherorder}
\begin{split}
           \partial_t h(s)   = &\partial_s^2 h(s)+ \int_0^{x} \upd y \, h(y) |s-y|^{-\lambda} \\  
      & +\frac{1}{2}  \partial_s^2 h(s) \int_0^{s} \upd y \, h(y) |s-y|^{2-\lambda}+ \xi(s) \, .
\end{split}
\end{equation}

The following results are applicable to multi-dimensional systems. We therefore now generalize to any spatial dimension by writing spatial coordinate in vector form, $\vec{s}$. In Fourier space, Eq.~\eqref{eq:langevin_higherorder} can then be written in compact form,
\begin{align}
\label{eq:long_distance_fourier}
    G_0 (\vec{q})^{-1} h (\vec{q},\omega)   = &  - \mu \int_{\vec{k}} W(\vec{q},\vec{k}) h(\vec{q},\omega) h(\vec{k-q},\omega) \\
    &+\xi(\vec{q},\omega)\, ,
\end{align}
where $h(\vec{q},t)$ is the Fourier transform of $h(s,t)$ and $\mu = \slfrac{(2-3 \lambda + \lambda^{2})}{2}$,
\begin{equation}
h(\vec{q},\omega) = \int \upd\vec{s} \int \upd t\, h\left(\vec{s},t\right) e^{i\vec{q}s}e^{i \omega t}\, .
\end{equation}
We also defined the free propagator,
\begin{equation}
        G_0^{-1} = \left( i\omega  + D\vec{q}^2 + J|\vec{q}|^{-\lambda}  \right)\, ,
\end{equation}
in which we reintroduced the dimensional parameters from the adimensional Eq.~\eqref{eq:long_distance_fourier} in order to study their flow under  renormalization. 
Finally, we defined the vertex, which accounts for the nonlinear terms, as
\begin{equation} 
 W(\vec{q},\vec{k}) =\frac{1}{2} \left[  \frac{\vec{q}(\vec{k-q})}{|\vec{k-q}|^{3-\lambda}}  + \frac{(\vec{k-q})\vec{q}}{|\vec{q}|^{3-\lambda}} \right]\, .
 \end{equation}
In the hydrodynamic limit, $\abs{\vec{k}} \rightarrow 0 $ and $\abs{\vec{q}}\to 0$, the vertex scales linearly with $\abs{\vec{k}}$ and $\abs{\vec{q}}$, which implies non-renormalization of the vertex function. 

The rescaling step of the renormalization group procedure then gives a rescaling of the form
\begin{equation}
    \begin{split}
    \partial_l \epsilon &= \left[z- 2 + A_{D} \right]\epsilon \, ,\\    
     \partial_l \mu &= \left[z + \chi -2  + (3-\lambda)\right]\mu\, , \\ 
      \partial_l \Gamma_C &= \left[z- 2\chi - d - 2 +A_{\Gamma_C} \right]\Gamma_C\, , \\
      \partial_l \Gamma_{NC} &= \left[z- 2\chi - d +A_{\Gamma_{NC}} \right]\Gamma_{NC}\, .
    \end{split}
 \end{equation}
 $A_{\Gamma_{NC}}$ and $A_{\Gamma_{C}}$ depend on all of the parameters and need to be evaluated perturbatively by calculating the diagrams in Fig.~\ref{fig:diagrams}.

Following standard calculations for the integrals represented by the diagrams in Fig.~\ref{fig:diagrams}~\cite{ZinnJustin}, we recover the renormalization-group flow for the parameters,
\begin{align}
    \partial_l \ln(\epsilon) =& z- 2 \\
    &- \frac{K_d\mu^2}{d \epsilon^3}  \left[\left(d-2  \right) \Gamma_{NC} + \left(d-3  \right)\Gamma_C\right] , \\   
     \partial_l \ln(\mu) =& z + \chi-2+(3-\lambda), \\ 
      \partial_l \ln(\Gamma_C)=& z- 2\chi - d - 2 \\
      &-\frac{K_d \nu^2(1+d) (\Gamma_{NC} +  \Gamma_C)^2}{2d \sigma_0^3\Gamma_C}  , \\
      \partial_l \ln(\Gamma_{NC}) = & z- 2\chi - d \, ,
 \end{align}
where $K_d = S_d/(2\pi)^d$ and $S_d$ is the surface area of a $d$-dimensional sphere.
 
From the non-renormalization of the non-conserved noise and of the couplings $\mu$ we get the exact exponent identities $\chi = \slfrac{(-1-d+\lambda)}{3}$ and $z = \slfrac{(-2+d+2\lambda)}{3}$, where $z$ is the dynamical critical exponent.  In $d=1$, two-point spatial correlations functions then decay with an exponent $\slfrac{2(\lambda-2)}{3}$ in the long-distance regime. 

\begin{figure}[t!]
    \centering
\includegraphics[width=0.95\linewidth]{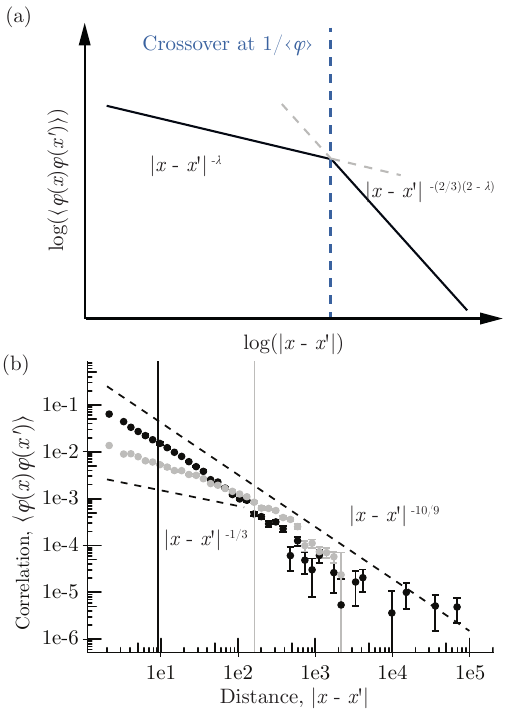}
    \caption{(a) The two-point correlation function decays in two spatial regimes with exponents given by Eq.~\eqref{eq:corr_function}. The position of the cross-over between both regimes increases with the typical length scale associated with the enzyme density, which scales as $1/\langle\phi\rangle$. (b) Numerical simulation of the spatial correlation functions for $\lambda = 1/3$ from stochastic simulations using Gillespie's algorithm with $10^7$ lattice sites. We calculated the correlation function in two stages of the simulation: for an average occupancy probability of 0.006 (grey, emphasizing the short-distance regime) and an average occupancy of 0.108 (black, emphasizing the long-distance regime). The vertical lines indicate the corresponding crossover positions. Dashed lines indicate the predicted exponents in both regimes. We binned the data logarithmically and error bars indicate mean $\pm$ standard error. }
\label{fig:PredictionCorrelationFunctions}
\end{figure}

Taken together, we find that the correlation function along one-dimensional substrates decays in two algebraic regimes [Fig.~\ref{fig:PredictionCorrelationFunctions}(a)],
\begin{align}
 \langle h(s)h(s')\rangle = 
 \begin{cases}
 |s-s'|^{-\lambda}. & \text{for } |s-s|\ll 1/\langle \sigma \rangle \, , \\
 |s-s'|^{-\frac{2}{3}(2 -\lambda)}, &\text{for } |s-s|\gg 1/\langle \sigma \rangle  \, .
 \end{cases} \label{eq:corr_function}
 \end{align}
The cross-over between these regimes stems from an effective exponential cutoff of the long-range interactions. The position of the cross-over scales with the only length scale in the system, the typical distance between neighboring occupied sites, $1/\langle \sigma \rangle$, and, intuitively, separates a regime dominated by long-range interactions and a regime characterized by passive, conservative fluctuations.

Figure~\ref{fig:diagrams}(b) shows correlation functions obtained from numerical simulation of Eq.~\eqref{eq:master_eq} for the special case of a space-filling substrate ($\lambda=1/3$). In order to separately emphasize the short-distance and the long-distance regime we computed correlations at two levels of the average substrate occupancy. While we cannot confidently estimate the numerical exponent for the short-distance regime, the simulation data confirm the existence of a crossover and the exponent of $-10/9$ in the long-distance regime.
 
\section{Discussion}
In summary, we studied a stochastic enzyme-substrate model where binding events are correlated via long-range interactions. Such interactions mimic the effect of conformational changes in the substrate, where positions far apart along the substrate might be close in physical space. We employed a coherent-state path-integral representation of the master equation and renormalization-group theory to calculate the exponents describing the time evolution of the average occupancy and the 

In polymer systems, the exponent $\lambda$ relates to the end-to-end distance of the polymer. For example, for a space-filling polymer, where $\lambda = 1/3$, the average occupancy increases with $t^{3/2}$ while the correlation function decays with $|x-x'|^{-1/3}$ for short distances and $|x-x'|^{-10/9}$ for long distances. Our results were obtained in a limit, where two-body interactions dominate. If substrate occupation is high, we expect that these exponents might be modified to incorporate higher-order interactions.

Novel technologies in super-resolution microscopy can quantify the location of individual enzymes bound to membranes or the DNA~\cite{Reinhardt:2023aa}. Such technologies allow for the quantification of spatial correlation functions of bound enzymes. Our work allows associating the spatio-temporal statistics of bound enzymes to the enzyme-substrate kinetics and spatial confirmation of the substrate. Measuring the moments along the substrate in an experiment would therefore allow drawing conclusions about the underlying biochemical processes.

\section{Acknowledgments}
We thank F. Piazza , M. Henkel, and F. J\"ulicher for helpful feedback and the entire Rulands group for fruitful discussions. We thank W. Reik, S. Clark, T. Lohoff, and I.  Kafetzopoulos for fruitful discussions about the biological aspects of this work. This project has received funding from the European Research Council (ERC) under the European Union’s Horizon 2020 research and innovation program (grant agreement no. 950349). This project has received funding from the European Union’s Horizon 2020 research and innovation programme under the Marie Skłodowska-Curie grant agreement No 101034413.

\appendix
\section{Path Integral Formulation of the Master Equation}
\label{sec:AppendixA}

To construct the field theory we first note that we can formally write a solution of the master equation as
\begin{equation}
\label{eq:exponentialmaster}
 |P(t) \rangle= e^{- Ht }|P(0) \rangle\, ,
\end{equation}
where $|P(0) \rangle$ is the initial state (i.e. the probability distribution of enzyme binding profiles at time $t=0$). The exponential factor can be rewritten as
\begin{equation}
e^{- H t } =(1-\Delta t H)^{\frac{t}{\Delta t}} = (1-\Delta t H) \cdot (1-\Delta t H) \cdot \ldots \, .
\end{equation}
By inserting the identity in the coherent state basis between every factor on the right-hand side of Eq.~\eqref{eq:exponentialmaster}, we find that the solution for any time $t_1$ can be written as

\begin{equation}
\begin{split}
&|P(t_1)\rangle  =\int \prod \sigma_i d\phi_i(t_1+\Delta t) d\hat{\phi}_i(t_1+\Delta t)  d\phi(t_1) d\hat{\phi}_i(t_1) \\
& \cdot e^{-\hat{\phi}_i(t_1)\phi(t_1)} \cdot e^{-\hat{\phi}_i(t_1 +\Delta t)\phi(t_1 +\Delta t)_i}
\cdot e^{\phi_i(t_1 +\Delta t)a^{\dagger}}\\
&\cdot\ket{0}\bra{0}e^{\hat{\phi}_i(t_1 +\Delta t) a}\left(1-\Delta t H\right) e^{\phi _i(t_1)a^{\dagger}}\ket{0}\bra{0}e^{\hat{\phi}_i(t_1)a} \, .
\end{split}
\end{equation}

In this equation, we have to evaluate quantities in the coherent state basis between the bra and the ket,
\begin{equation}
\begin{split}
 & \bra{0}e^{\hat{\phi_i}(t_1 +\Delta t) a}\left(1-\Delta t H\right) e^{\phi_i (t_1)a^{\dagger}}\ket{0} =  \\
 & e^{\hat{\phi_i}(t_1 +\Delta t) } e^{\phi_i(t_1 ) } -\Delta  t \bra{0}e^{\hat{\phi_i}(t_1 +\Delta t) a}\left(H\right) e^{\phi _i(t_1)a^{\dagger}}\ket{0} \approx \\
& e^{\hat{\phi_i}(t_1 +\Delta t) } e^{\phi_i(t_1 ) }e^{-\Delta t H(\hat{\phi_i}(t_1),\phi_i(t_1))}\, ,
\end{split}
\end{equation}
where $H(\hat{\phi}_i,\phi_i)$ is obtained by replacing all $a_i$ with $\phi_i $ and $a^{\dagger}_i$ with $\hat{\phi}_i$, using the previously introduced decomposition and by taking into account the rules for exclusion processes, Eq.\eqref{eq:exclusion}. Repeating this procedure $\slfrac{t}{\Delta t }$ times for each factor $(1-\Delta t H)$ we end up with an integral, over a product of three terms $P_2$, $P_3$, $P_4$. The integral is,
\begin{equation}
 \int \prod\sigma_i d\hat{\phi}_i(t) d\phi_i(t )  \ldots d\hat{\phi}_i(\Delta t) d\phi_i(\Delta t) d\hat{\phi}_i(0) d\phi_i(0)\, ,
\end{equation}
which can be rewritten compactly as a functional integral $\int \mathcal{D}[\phi]\mathcal{D} [\hat\phi] $.
$P_2$ is composed of a product of terms that can be rewritten utilizing Riemann integration,
\begin{equation}
P_2 = \prod_{t_1 = \Delta t} ^{t} e^{\hat{\phi}(t_1 +\Delta t) \phi(t_1 )  - \hat{\phi}(t_1) \phi(t_1 ) } \approx e^{-\int dt\partial_t \hat{\phi} \phi}.
\end{equation}
Finally, there are further $\slfrac{t}{\Delta t}$ terms coming from the Hamiltonian evaluated at each time step,
 \begin{equation}
P_3 = \prod_{t_1 = \Delta t} ^{t} e^{-\Delta t H(\hat{\phi}(t_1),\phi(t_1))} \approx e^{-\int dt  H(\hat{\phi}(t),\phi(t))}\, . 
\end{equation}
The final factor, $P_4$, represents initial conditions and we refer to Ref.~\cite{Cardy} for a discussion of this term. With this, any observable can then be expressed as a path integral of the form
\begin{equation}
A(\vec{\sigma}) = \int \mathcal{D}[\phi]\mathcal{D} [\hat\phi] A(\boldsymbol{\phi},\boldsymbol{\hat{\phi}}=1) e^{-S[\hat{\phi},\phi] }\, ,
\end{equation}
with
\begin{equation}
S[\hat{\phi},\phi] = - \sum_{i}\phi_{i}(t_f) + \int_{0}^{t_f} \mathrm{d}t \sum_{i}\left(\hat{\phi}_{i}(t) \partial_t \phi_i(t) 
+ H_i[\hat{\phi},\phi]\right)\, ,
\end{equation}
where we have performed a partial integration in time and,
\begin{equation}
\begin{split}
    \label{hamiltonian1}
&H_i[\hat{\phi},\phi] =(1-\hat{\phi}_i)  e^{-\hat{\phi}_i \phi_i} \\
&\left( \sum_{l=1}^{N-i}\prod_{j=1}^{l-1}  J_{i,i+l}(l) \hat{\phi}_{i+l}  \phi_{i+l} 
(1- \phi_{i+j}) + +\textit{o.t.}\right)\, .
\end{split}
\end{equation}
Defining the generating functional of correlations, $Z[\vec{h},\phi,\hat{\phi}]$, as
\begin{equation}
\label{correlationFT}
Z[\vec{h},\phi,\hat{\phi}]= \int \mathcal{D}[\vec{h},\phi,\hat{\phi}]   e^{-S[\phi,\hat{\phi}] +\int \int ds dt \left[h \phi + \hat{h}\hat{\phi} \right]}\, ,
\end{equation}
expectation values of products of observables, such as correlation functions, can then be expressed as functional derivatives with respect to the auxiliary external field,
\begin{equation}
\langle \phi (s,t)  \phi (y,t') \rangle = \frac{\delta^2}{\delta h(s,t) \delta h(y,t)} Z[\vec{h},\phi,\hat{\phi}]|_{\vec{h}=0} \, .
\end{equation}

\section{Stochastic simulations}
\label{sec:numericalsimul}
To test the validity of our analytical results we perform extensive stochastic simulations by integration of the master equation Eq.~\eqref{eq:master_eq} using Gillespie's algorithm \cite{Gillespie} with  $1e7$ lattice sites, a random initial distribution of enzymes with a occupancy fraction equal to $10^{-4}$, and interaction strength $J=1$. Unbinding and hopping rates are set to zero as they do not affect the critical dynamics and would lead to significantly slower simulations. This is because the number of lattice sites needs to be large enough to measure the exponents of the spatial correlations.

\bibliography{Biblio}

 \end{document}